\documentclass[letterpaper]{article} 
\usepackage{aaai2026}  
\usepackage{times}  
\usepackage{helvet}  
\usepackage{courier}  
\usepackage[hyphens]{url}  
\usepackage{graphicx} 
\usepackage{natbib}  
\usepackage{caption} 
\usepackage{algorithm}
\usepackage{algorithmic}

\usepackage{subcaption}
\usepackage{enumitem}
\usepackage{multicol}
\usepackage{multirow}
\usepackage{makecell}
\usepackage{bm}
\usepackage{xcolor}
\usepackage{colortbl}
\definecolor{my_gray}{rgb}{0.91, 0.91, 0.91}
\definecolor{my_green}{rgb}{0.078, 0.686, 0.384}
\definecolor{my_purple}{rgb}{0.858, 0.851, 0.894}
\definecolor{my_gray_light}{rgb}{0.949, 0.949, 0.949}

\usepackage{newfloat}
\usepackage{listings}
\DeclareCaptionStyle{ruled}{labelfont=normalfont,labelsep=colon,strut=off} 
\floatstyle{ruled}
\newfloat{listing}{tb}{lst}{}
\floatname{listing}{Listing}
\title{PEOAT: Personalization-Guided Evolutionary Question \\Assembly for One-Shot Adaptive Testing}

\author{
    Xiaoshan Yu\textsuperscript{\rm 1},
    Ziwei Huang\textsuperscript{\rm 2},
    Shangshang Yang\textsuperscript{\rm 3}\thanks{Corresponding author.},
    Ziwen Wang\textsuperscript{\rm 3},\\
    Haiping Ma\textsuperscript{\rm 4},
    Xingyi Zhang\textsuperscript{\rm 3}
}
\affiliations{
    \textsuperscript{\rm 1}School of Artificial Intelligence, Anhui University, China\\
    \textsuperscript{\rm 2}School of Computing and Information Systems, Singapore Management University, Singapore\\
    \textsuperscript{\rm 3}School of Computer Science and Technology, Anhui University, China\\
    \textsuperscript{\rm 4}Institutes of Physical Science and Information Technology, Anhui University, China\\
    \{yxsleo, yangshang0308, wzw12sir, xyzhanghust\}@gmail.com, ziweihuang@smu.edu.sg, hpma@ahu.edu.cn
}

\usepackage{enumitem}
\usepackage{booktabs}
\usepackage{bm}
\usepackage{multirow}
\usepackage{xcolor}
\usepackage{amsmath}
\usepackage{amsfonts}

\begin{document}

\maketitle

\begin{abstract}

With the rapid advancement of intelligent education, Computerized Adaptive Testing (CAT) has attracted increasing attention by integrating educational psychology with deep learning technologies. Unlike traditional paper-and-pencil testing, CAT aims to efficiently and accurately assess examinee abilities by adaptively selecting the most suitable items during the assessment process. However, its real-time and sequential nature presents limitations in practical scenarios, particularly in large-scale assessments where interaction costs are high, or in sensitive domains such as psychological evaluations where minimizing noise and interference is essential. These challenges constrain the applicability of conventional CAT methods in time-sensitive or resource-constrained environments. To this end, we first introduce a novel task called one-shot adaptive testing (OAT), which aims to select a fixed set of optimal items for each test-taker in a one-time selection. Meanwhile, we propose \textbf{PEOAT}, a \underline{\textbf{P}}ersonalization-guided \underline{\textbf{E}}volutionary question assembly framework for \underline{\textbf{O}}ne-shot \underline{\textbf{A}}daptive \underline{\textbf{T}}esting from the perspective of combinatorial optimization. Specifically, we began by designing a personalization-aware initialization strategy that integrates differences between examinee ability and exercise difficulty, using multi-strategy sampling to construct a diverse and informative initial population. Building on this, we proposed a cognitive-enhanced evolutionary framework incorporating schema-preserving crossover and cognitively guided mutation to enable efficient exploration through informative signals. To maintain diversity without compromising fitness, we further introduced a diversity-aware environmental selection mechanism. The effectiveness of PEOAT is validated through extensive experiments on two datasets, complemented by case studies that uncovered valuable insights.

\end{abstract}


\section{Introduction}

Computerized adaptive testing~(CAT)~\cite{wainer2000computerized}, as a significant and promising approach to personalized assessment in intelligent education~\cite{AIED_1,AIED_3}, has garnered increasing attention and development in recent years. Its goal is to deliver adaptive ability evaluation for students through progressive interaction and feedback, integrating principles from educational psychology with advances in deep learning. In general, CAT consists of two primary modules~\cite{liu2024survey,MKLI}: a question selection module $\mathcal{M}_{\pi}$ and a cognitive diagnosis module $\mathcal{M}_d$. The former adaptively selects the most suitable questions based on the test taker's current ability status, while the latter diagnoses the test taker's knowledge proficiency based on response feedback, as illustrated in Figure~1(a). These two modules operate alternately until a predefined termination condition is satisfied.

\begin{figure}[!t] 
    \centering 
    \includegraphics[width=0.97\linewidth]{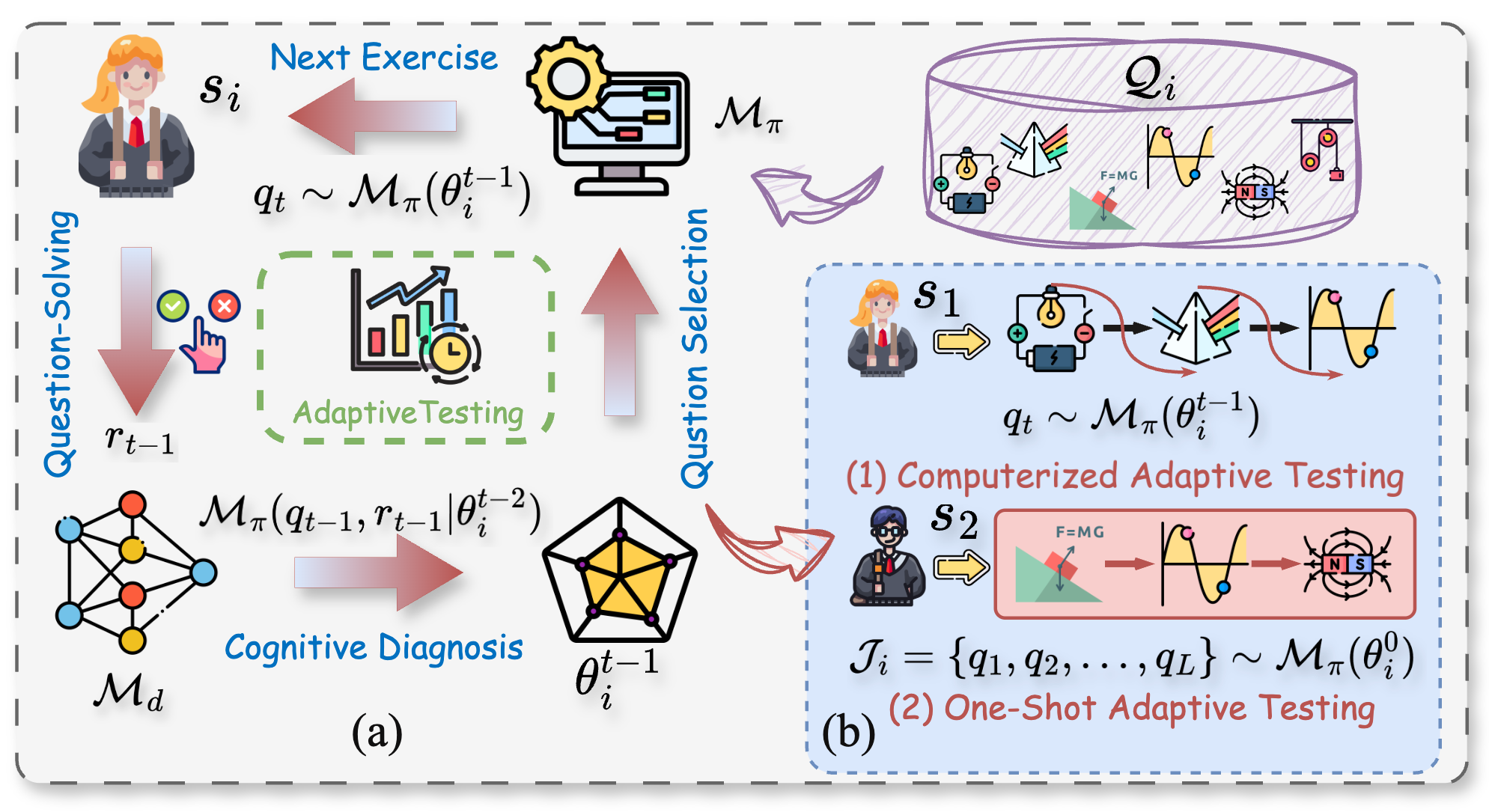} 
    \vspace{-3mm}
    \caption{(a)~The process of computerized adaptive testing; (b)~Comparison of two adaptive testing tasks~(CAT \& OAT).}
    \vspace{-4mm}
    \label{fig.gcd} 
\end{figure}

Existing research on CAT~\cite{liu2024survey} primarily focuses on enhancing the question selection algorithm, which are widely regarded as key determinants of assessment adaptability and effectiveness. These approaches can be broadly categorized into heuristic methods and data-driven learning methods. Heuristic approaches~\cite{MAAT,BECAT} rely on explicitly defined, interpretable rules to select items that align question characteristics with the test taker’s estimated ability. For instance, BECAT~\cite{BECAT} approximates full-response gradients to guide item selection, enabling accurate ability estimation with fewer questions and offering theoretical guarantees on estimation error. In contrast, data-driven methods~\cite{BOBCAT,NCAT} seek to improve performance by learning personalized item selection policies directly from data. A representative example is NCAT~\cite{NCAT}, which views CAT as a bilevel reinforcement learning problem, where an attentive policy is trained to select items by modeling learning behavior.

Despite existing CAT methods have demonstrated notable success, their inherent interactivity, requiring iterative item selection and ability estimation, poses significant limitations in scenarios with high interaction costs or constrained response conditions. In many real-world scenarios, such as psychological assessments~\cite{psychological}, post-instruction diagnostic evaluations~\cite{post-instruction}, or remote/offline testing~\cite{offline}, the feasibility of interactive testing is often hindered by factors such as response latency, user anxiety, or device limitations. To address this gap, this paper proposes a novel task called One-Shot Adaptive Testing (OAT), in which a fixed set of candidate items is adaptively selected beforehand and presented to the test-taker all at once, as illustrated in Figure 1(b). Considering the characteristics of this problem, we attempt to model it from a combinatorial optimization perspective.

However, this task is challenging mainly due to three issues: (1) \textit{ensuring student adaptability during optimization}; (2) \textit{searching effectively in a vast solution space}; (3) \textit{mitigating encoding sparsity given a candidate pool much larger than the test length}. To this end, in this paper, we propose \textbf{PEOAT}, a \underline{\textbf{P}}ersonalization-guided \underline{\textbf{E}}volutionary question assembly framework for \underline{\textbf{O}}ne-hot \underline{\textbf{A}}daptive \underline{\textbf{T}}esting. Specifically, we first propose a personalization aware-based population initialization strategy that accounts for individual student ability differences and exercise difficulty, employing multi-strategy sampling to generate a diverse initial question population and effectively construct the initial search space. Next, we develop a cognitive-enhanced evolutionary search framework, featuring the schema-preserving uniform crossover and the cognitive information-guided mutation operators that leverage informative cues throughout population evolution for efficient exploration. Finally, we design a diversity-preserving environmental selection strategy that balances diversity maintenance with fitness during offspring selection. Extensive experiments on two real-world educational datasets validate the effectiveness of the proposed PEOAT model. Additionally, we conduct insightful case studies that reveal valuable findings.

\section{Related Work}

\subsection{Computerized Adaptive Testing}

As a core assessment paradigm in  personalized education, computerized adaptive testing~(CAT)~\cite{wainer2000computerized} originated from educational psychology and has evolved through the incorporation of deep learning techniques~\cite{HD-KT,HashCAT,li2025pcat}. It aims to achieve accurate ability diagnosis by interactively selecting suitable exercises in response to test-taker performance. Recent advances in CAT have predominantly focused on improving item selection strategies, generally falling into two categories~\cite{MKLI,liu2024survey,yu2024rigl}: heuristic methods and data-driven approaches. The former~\cite{KLI,MKLI,BECAT,MAAT,ma2025learning,yang2024endowing} selects questions based on explicitly defined and interpretable rules, aiming to match question characteristics with the test-taker’s estimated ability. For example, Maximum Fisher Information~(MFI)~\cite{MFI} minimizes ability estimation variance via local item information, whereas KLI~\cite{KLI} improves robustness by incorporating global Kullback-Leibler divergence. Moreover, MAAT~\cite{MAAT} defines the informativeness of exercises based on the expected maximum change criterion from active learning. In contrast, data-driven methods~\cite{BOBCAT,NCAT,GMOCAT,UATS} aim to enhance performance by learning personalized selection policies directly from learner-exercise interaction data. Representatively, NCAT~\cite{NCAT} casts CAT as a bilevel reinforcement learning problem, where an attentive neural policy is trained to select items by directly modeling student behaviors~\cite{gao2025agent4edu,gao2024zero,xiaoshan2024disco}. Although these methods have achieved notable success, they are often impractical in resource-constrained ability assessment scenarios, highlighting the need for one-shot adaptive testing, which serves as the primary motivation for this study.

\subsection{Evolutionary Optimization Application}

Combinatorial optimization~\cite{papadimitriou1998combinatorial,blum2003metaheuristics} refers to the process of searching for an optimal object from a finite but often exponentially large solution space, and it plays a central role in various complex decision-making tasks~\cite{LIGHT,yang2025fedcd,haiping2024dgcd}. When the solution space lacks closed-form structure or involves complex constraints, gradient-based methods~\cite{lezcano2019trivializations} often fail, making heuristic strategies, particularly evolutionary algorithms~(EAs), a compelling alternative~\cite{yang2023evolutionary,yu2024rdgt}. Over the past decades, a wide variety of evolutionary algorithms~\cite{vcrepinvsek2013exploration} have been proposed and refined. Classical examples include the Genetic Algorithm~(GA)~\cite{lambora2019genetic}, which mimics natural selection through genetic operators, and Differential Evolution~(DE)~\cite{das2010differential}, which leverages vector-based mutations for continuous and combinatorial tasks. These methods have proven effective in various domains and are gaining increasing traction in education~\cite{yang2023cognitive,bu2022cognitive,sun2022genetic,bu2023probabilistic}, where they are used to tackle complex decision-making problems. For example, PEGA~\cite{yang2023cognitive} employs a constrained multi-objective framework with dual co-evolution to assemble personalized exercise groups~\cite{liu2023homogeneous,yu2024rdgt}. In the cognitive diagnosis~\cite{yang2025disengcd,dong2025knowledge}, HGA-CDM~\cite{bu2022cognitive} applies a memetic algorithm combining genetic and adaptive local search to the DINA model, mitigating its exponential computational complexity. However, how to effectively model the OAT task from an evolutionary optimization perspective remains unexplored and presents a valuable research direction.


\begin{figure*}[!t] 
	\centering 
        \vspace{-2mm}
	\includegraphics[scale=0.201]{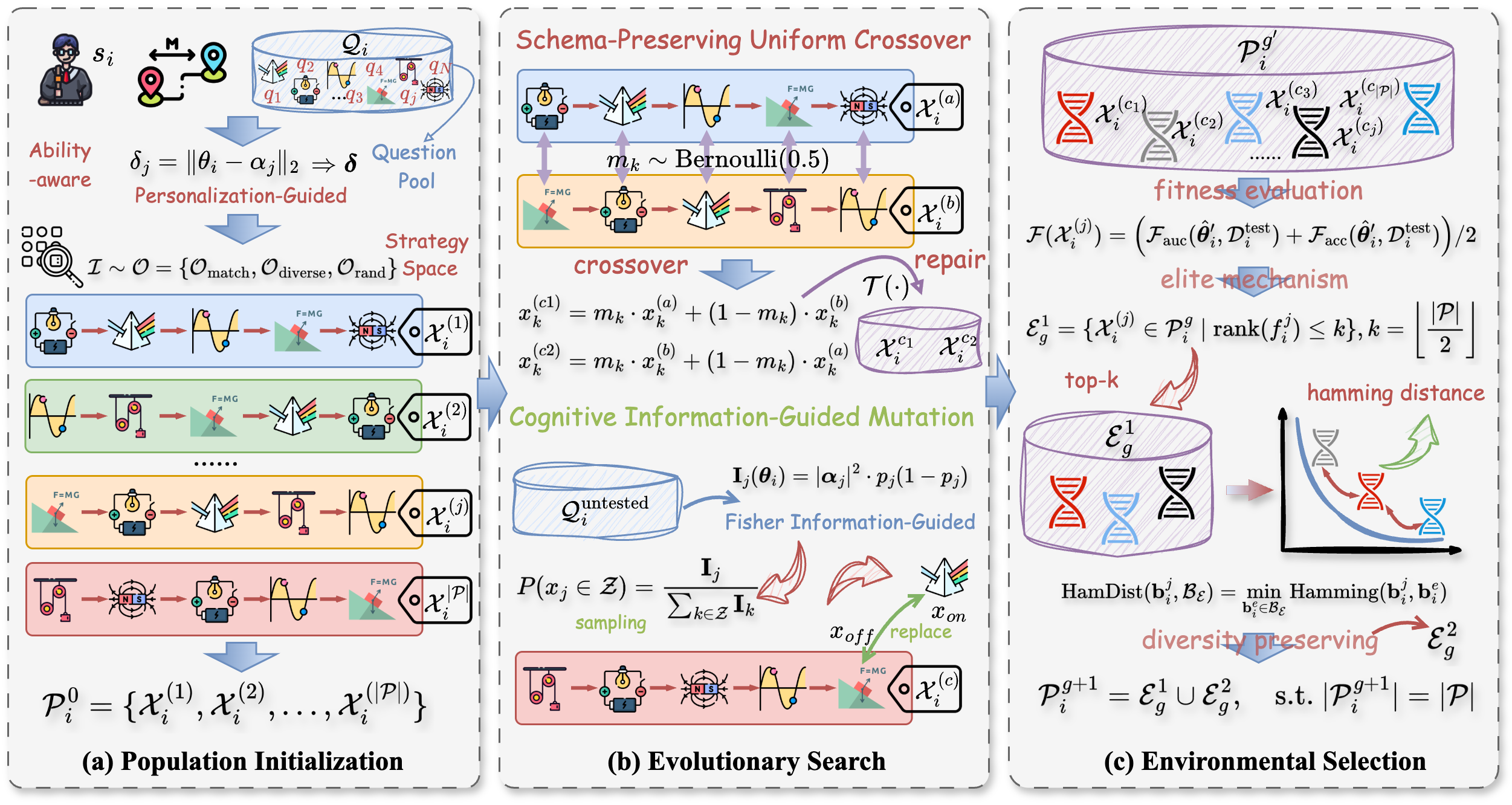} 
    \vspace{-2mm}
    \caption{The overview architecture of our proposed PEOAT model. (a) The personalization-aware population initialization. (b) The the cognitive-enhanced evolutionary search. (c) The diversity-preserving environmental selection. Best viewed in color.}
        \label{fig.framework} 
        \vspace{-5mm}
\end{figure*}

\section{Preliminary}

\subsection{Problem Statement}
In this section, we formally define the One-Shot Adaptive Testing~(OAT) task. In an intelligent education system, let $\mathcal{S}=\{s_1,s_2,\ldots,s_N\}$ be the set of $N$ students, $\mathcal{Q}=\{q_1,q_2,\ldots,q_M\}$ be the candidate pool of $M$ questions, and $\mathcal{C}=\{c_1,c_2,\ldots,c_K\}$ be the set of $K$ knowledge concepts. The mapping between questions and knowledge concepts is commonly represented by a $Q$-matrix, denoted as $\mathbf{Q} = \{m_{ij}\}^{M \times K}$. In this matrix, an entry $m_{ij} = 1$ signifies that question $e_i$ is linked to concept $c_j$, while $m_{ij} = 0$ indicates no such association. For each student $s_i \in \mathcal{S}$ with historical assessment records, their interactions can be represented as $\mathcal{R}_i = \{(s_i, q_j, r_{ij}) \mid q_j \in \mathcal{Q},\ r_{ij} \in \{0,1\}\}$, where $r_{ij} = 1$ denotes a correct response to question $q_j$, and $r_{ij} = 0$ otherwise. The complete \textbf{\textit{One-Shot Adaptive Testing~(OAT)}} system is composed of two fundamental components: \textbf{(1)} the \textit{cognitive diagnosis module} $\mathcal{M}_d$ that models the examinee’s knowledge proficiency by predicting the probability of correctly answering each question $q$~\cite{gao2024collaborative,yu2025rethinking}, denoted as $\mathcal{M}_d(q \mid \theta) \in [0, 1]$; and \textbf{(2)} the \textit{question selection module} $\mathcal{M}_\pi$ that selects a subset of $L$ questions $\mathcal{J} \subset \mathcal{Q}$ in a one-shot manner, based on an initial ability $\theta^{0}$. More specifically, given the initial ability estimate $\theta_i^0$ of examinee $s_i$, the OAT selects a fixed-length question set $\mathcal{J}_i=\{q_1,q_2,\ldots,q_L\}\sim \mathcal{M}_\pi (\theta_i^0)$ without any intermediate feedback during the test process. After the examinee finishes all $L$ questions and their responses $\mathbf{r} = \{r_{i1}, r_{i2}, \ldots, r_{iL}\}$ are collected, the diagnostic model $\mathcal{M}_d$ conducts a single-step ability update to produce the final proficiency estimate $\theta_i^{\text{final}}$. In contrast to conventional CAT, where questions are selected sequentially as $q_t\sim \mathcal{M}_\pi(\theta_i^{t-1})$ and ability estimates $\theta_i^t$ are updated iteratively after each response, OAT aims to estimate the true knowledge proficiency $\hat{\theta}_i$ as accurately and efficiently as possible using only a single batch of adaptively selected questions, i.e., $\theta_i^{\text{final}} \rightarrow \hat{\theta}_i$.

\subsection{Combinatorial Optimization Perspective}

Unlike CAT, which selects questions in a sequential and feedback-driven manner~\cite{UATS}, OAT poses a distinct challenge: selecting an optimal fixed-length question set in a single round without any intermediate feedback. This constraint requires the selection policy to holistically consider the test-taker's prior ability and question characteristics to maximize the diagnostic utility of the selected items. From a modeling perspective, this task can be naturally viewed as a bi-level combinatorial optimization problem, where the outer layer selects a subset of questions, and the inner layer estimates student ability based on simulated response data. The optimization objective is to ensure that the final ability estimation is as close as possible to the student's true proficiency. 
Formally, from the perspective of discrete combinatorial optimization, the OAT task for each student $s_i\in \mathcal{S}$ can be characterized as follows:
\begin{equation}
    \setlength{\abovedisplayskip}{3pt}
    \setlength{\belowdisplayskip}{3pt}
    \label{oat_obj}
    \left\{
    \begin{aligned}
        & \mathcal{J}_i^* = \mathop{\arg\max}\limits_{\mathcal{J}_i \subseteq \mathcal{Q}_i^{\text{untested}}} \mathcal{F}(\theta_i^{\text{final}}(\mathcal{J}_i), \;\hat{\theta}_i), \\
        & \text{s.t.} \;\; \theta_i^{\text{final}}(\mathcal{J}_i) = \mathop{\arg\min}\limits_{\theta_i} \sum_{(q_j, r_{ij})}^{\mathcal{R}_i(\mathcal{J}_i)} \mathcal{L}\left(r_{ij}, \mathcal{M}_d(q_j \mid \theta_i)\right), \\
        & \text{where} \;\; \mathcal{J}_i\sim \mathcal{M}_\pi (\theta_i^0), \quad \text{and} \\
        & \quad\quad\quad \mathcal{R}_i(\mathcal{J}_i)=\{(q_1,r_{i1}),\ldots,(q_L,r_{iL}) \mid q_i \in \mathcal{J}_i))\}.
    \end{aligned}
    \right.
\end{equation}


\section{Methodology}
In this section, we present the PEOAT framework in detail. As depicted in Figure~2, the PEOAT is composed of three key components: the personalization-aware population initialization, the cognitive-enhanced evolutionary search, and the diversity-preserving environmental selection.

\subsection{Personalization-Aware Population Initialization}

To effectively guide the evolutionary search in OAT, we design a personalization-aware population initialization mechanism that adaptively generates an informative and diverse initial population based on students' personal abilities and the characteristics of the candidate exercises. As mentioned earlier, the one-shot item selection process for each student can be modeled as a population-based evolutionary optimization procedure. Accordingly, for each student $s_i\in\mathcal{S}$, every individual in the population represents a candidate test form consisting $L$ questions, encoded as follows: 
\begin{equation}
    \label{eq_2}
    \begin{split}
        \mathcal{X}_{i}^{(j)} = [x_{1}, x_{2}, \ldots, x_{L}] \in {\mathcal{Q}_i^{\text{untested}}}^L,
    \end{split}
\end{equation}
where $\mathcal{X}_{i}^{(j)}$ denotes the $j$-th individual in the candidate population of student $s_i$, and $x_{k}$ represents the $k$-th gene in the chromosome, which indexes a question from the student's remaining question pool, i.e., $x_{k} \rightarrow q_{k} \in \mathcal{Q}_i^{\text{untested}}$, and $\mathcal{Q}_i^{\text{untested}}$ denotes the set of untested questions for student $s_i$. We assume that all selected indices are unique—i.e., $q_{k_1} \ne q_{k_2}$ for $k_1 \ne k_2$—thereby satisfying the fixed-length constraint $|\mathcal{X}| = L$. This subset-based encoding not only defines the structure of each individual but also serves as a retrieval mechanism for latent features (e.g., question embeddings or difficulty parameters). Compared to sparse one-hot encodings, it offers a more compact and efficient representation, particularly suited for large-scale optimization.

To embed personalized prior knowledge into the search space while effectively balancing exploitation and exploration, we propose a multi-strategy population initialization mechanism. Specifically, we define a strategy space $\mathcal{O} = \{\mathcal{O}_{\text{match}}, \mathcal{O}_{\text{diverse}}, \mathcal{O}_{\text{rand}}\}$, representing three initialization strategies that select candidate exercises based on students’ initial abilities: matching, diverse, and random, respectively—each encouraging a distinct form of exploration. For each individual, one strategy is randomly sampled from $\mathcal{O}$, and the process of constructing question index gene-encoded candidates can be formalized as follows:
\begin{equation}
    \small
    \label{eq_3}
    \mathcal{I}_i \sim \left\{
    \begin{aligned}
        &\big\{\text{Uniform}( \text{Top}_{2L}(\boldsymbol{\delta}_i^{\uparrow}))\big\}^L,\quad \text{if } \mathcal{O}^{'} = \mathcal{O}_{\text{match}}, \\
        &\big\{\text{Uniform}( \text{Top}_{2L}(\boldsymbol{\delta}_i^\downarrow))\big\}^L,\quad \text{if } \mathcal{O}^{'} = \mathcal{O}_{\text{diverse}}, \\
        &\big\{\text{Uniform}(\boldsymbol{\delta}_i^\rightarrow[2L:-2L])\big\}^L,\quad \text{if } \mathcal{O}^{'} = \mathcal{O}_{\text{rand}},
    \end{aligned}
    \right.
\end{equation}
where $|\mathcal{I}| = L$, and $\boldsymbol{\delta}_i^{\uparrow}$ and $\boldsymbol{\delta}_i^{\downarrow}$ denote the ascending and descending sorted indices of $\boldsymbol{\delta}_i$, respectively. The vector $\boldsymbol{\delta}_i = [\delta_1, \delta_2, \ldots, \delta_{|\mathcal{Q}_i|}]$ represents a personalized distance vector that quantifies the matching quality between student $s_i$ and the questions in $\mathcal{Q}_i$, where each $\delta_j$ is computed as:
\begin{equation}
    \label{eq_4}
    \begin{split}
        \delta_j = \|\boldsymbol{\theta}_i-\boldsymbol{\alpha}_j\|_2,\quad \forall j\in\{1,2,\ldots,|\mathcal{Q}_i|\},
    \end{split}
\end{equation}
where $\boldsymbol{\theta}_i$ and $\boldsymbol{\alpha}_j$ represents the ability vector of student $s_i$ and the difficulty vector of question $q_j$, respectively. Each resulting index set $\mathcal{I}_i$ is subsequently transformed into the corresponding individual encoding, i.e., $\mathcal{I}_i \rightarrow \mathcal{X}_i$. The final initialized population of predefined size $|\mathcal{P}|$ is given by:
\begin{equation}
    \label{eq_5}
    \begin{split}
        \mathcal{P}_i^0=\{\mathcal{X}_i^{(1)},\mathcal{X}_i^{(2)},\ldots,\mathcal{X}_i^{(|\mathcal{P}|)}\}.
    \end{split}
\end{equation}

\subsection{Cognitive-Enhanced Evolutionary Search}

To evolve high-quality question subsets tailored to individual examinees, we propose a cognitive-enhanced evolutionary search framework comprising two key operators: the schema-preserving uniform crossover operator and the cognitive information-guided mutation operator. Both operators maintain the fixed-length structure of individuals while being guided by the cognitive relevance signals.

\subsubsection{Schema-Preserving Uniform Crossover}

Let two parent individuals be denoted as $\mathcal{X}_i^{(a)} = [x^{(a)}_1, x^{(a)}_2, \ldots, x^{(a)}_L]$ and $\mathcal{X}_i^{(b)} = [x^{(b)}_1, x^{(b)}_2, \ldots, x^{(b)}_L]$, each representing a candidate question list. To generate two offspring $\mathcal{X}_i^{(c1)}$ and $\mathcal{X}_i^{(c2)}$, we sample a binary mask vector $\mathbf{m} \in \{0,1\}^L$ with $m_k \sim \text{Bernoulli}(0.5)$, and perform crossover as follows:
\begin{equation}
    \label{eq_6}
    \left\{
    \begin{aligned}
        x_k^{(c1)} &= m_k \cdot x_k^{(a)} + (1 - m_k) \cdot x_k^{(b)}, \\
        x_k^{(c2)} &= m_k \cdot x_k^{(b)} + (1 - m_k) \cdot x_k^{(a)},
    \end{aligned}
    \right.
\end{equation}
where $1\leq k \leq L$ denotes the crossover index, and the operator preserves individual structure while enabling fine-grained recombination, outperforming one-point or multi-point crossover in maintaining feasibility and diversity. To ensure that both offspring preserve uniqueness and validity (i.e., no duplicate questions and $\mathcal{X}_i^{(c)} \subset \mathcal{Q}_i^{\text{untested}}$), we apply a repair operator $\mathcal{T}(\cdot)$ that resolves duplicates by replacing them with randomly sampled non-overlapping items from the untested pool. The final offspring are given by:
\begin{equation}
    \label{eq_7}
    \begin{aligned}
        \mathcal{X}_i^{(c1)} \leftarrow \mathcal{T}\left(\mathcal{X}_i^{(c1)}\right), \quad \mathcal{X}_i^{(c2)} \leftarrow \mathcal{T}\left(\mathcal{X}_i^{(c2)}\right).
    \end{aligned}
\end{equation}

\subsubsection{Cognitive Information-Guided Mutation}

To introduce adaptive perturbation, we propose a mutation strategy that leverages personalized item information gain. For a given individual $\mathcal{X}_i = [x_1, \ldots, x_L]$, we randomly select a gene $x_{\text{off}}$ to remove, and then sample a replacement $x_{\text{on}}$ from the unselected pool based on an information-based distribution. Specifically, let $\boldsymbol{\theta}_i \in \mathbb{R}^d$ denote the ability vector of examinee $s_i$, and let $\boldsymbol{\alpha}_j \in \mathbb{R}^d$ be the difficulty vector of item $q_j$. According to the item response theory (IRT)~\cite{MIRT},  the probability that $s_i$ correctly answers $q_j$ is computed as: $p_j = \sigma(\boldsymbol{\theta}_i^\top \boldsymbol{\alpha}_j)$, where $\sigma(\cdot) = \frac{1}{1 + e^{-(\cdot)}}$ denotes the sigmoid function. To quantify how informative item $q_j$ is for estimating $\boldsymbol{\theta}_i$, we refer to the Fisher information matrix~\cite{Fisher}, which characterizes the expected curvature of the log-likelihood with respect to $\boldsymbol{\theta}_i$, and is defined as:
\begin{equation}
    \small
    \label{eq_8}
    \begin{aligned}
    \mathbf{I}_j(\boldsymbol{\theta_i})&=\mathbb{E}\big[\big(\frac{\partial}{\partial \boldsymbol{\theta}} \log p_j(\boldsymbol{\theta})^{r_{ij}} (1 - p_j(\boldsymbol{\theta}))^{1 - r_{ij}}\big) \left( \cdot \right)^\top \big],\\
    & = p_j (1 - p_j) \cdot \boldsymbol{\alpha}_j \boldsymbol{\alpha}_j^\top \in \mathbb{R}^{d \times d}, \quad \forall j \in \mathcal{Q}_i^{\text{untested}}.
    \end{aligned}
\end{equation}

However, directly manipulating this matrix in the mutation operator is computationally inefficient, especially when comparing information across many candidate items. To address this, we approximate the information matrix using its Frobenius norm~\cite{Frobenius} as a scalar proxy, yielding the scalar information gain for item $q_j$ as follows:
\begin{equation}
    \small
    \label{eq_9}
    \left\{
    \begin{aligned}
    \|\mathbf{I}_j(\boldsymbol{\theta}_i)\|_F &= p_j (1 - p_j) \cdot \|\boldsymbol{\alpha}_j \boldsymbol{\alpha}_j^\top\|_F = p_j (1 - p_j) \cdot \|\boldsymbol{\alpha}_j\|^2, \\
    \Rightarrow \mathbf{I}_j(\boldsymbol{\theta}_i) &= |\boldsymbol{\alpha}_j|^2 \cdot p_j (1 - p_j), \quad \forall j \in \mathcal{Q}_i^{\text{untested}}.
    \end{aligned}
    \right.
\end{equation}

Let $\mathcal{Z} = \mathcal{Q}_i^{\text{untested}} \setminus \mathcal{X}_i$ denote the pool of unselected candidate questions. We define a categorical sampling distribution over $\mathcal{Z}$ based on normalized information gain:
\begin{equation}
    \label{eq_10}
    \begin{aligned}
        P(x_j \in \mathcal{Z}) = \frac{\mathbf{I}_j}{\sum_{k \in \mathcal{Z}} \mathbf{I}_k},
    \end{aligned}
\end{equation}
where the new gene $x_{\text{on}}$ is then sampled from this distribution to replace the removed gene $x_{\text{off}}$, introducing a personalized, cognitively-informed mutation step that promotes high-information test composition. This mutation operator ensures that inserted genes are both personalized and cognitively informative, leading to more effective evolution.

\subsection{Diversity-Preserving Environmental Selection}

To ensure robust convergence and mitigate premature stagnation, we adopt a diversity-preserving environmental selection strategy. This mechanism balances fitness-oriented exploitation with diversity-aware exploration, ultimately forming the next-generation population with both high-quality and semantically diverse candidate question lists.

For each individual $\mathcal{X}_i^{(j)} = [x_{1}, x_{2}, \ldots, x_{L}]$, its fitness is assessed by simulating the one-shot assessment process. Specifically, student $s_i$ first completes the selected set of questions, after which the cognitive diagnosis model $\mathcal{M}_d$ performs a virtual parameter update to estimate the personalized knowledge ability vector, following the trajectory $\hat{\boldsymbol{\theta}}_i^0 \xrightarrow{\text{update}} \hat{\boldsymbol{\theta}}_i'$. The updated ability $\hat{\boldsymbol{\theta}}_i'$ is then evaluated on the reserved test set $\mathcal{D}_i^{\text{test}}$, and the prediction quality is measured using a hybrid metric that combines AUC and accuracy:
\begin{equation}
    \label{eq_11}
    \begin{aligned}
    \mathcal{F}(\mathcal{X}_i^{(j)}) = \left( \mathcal{F}_{\text{auc}}(\hat{\boldsymbol{\theta}}_i',\mathcal{D}_i^{\text{test}}) + \mathcal{F}_{\text{acc}}(\hat{\boldsymbol{\theta}}_i',\mathcal{D}_i^{\text{test}}) \right)/2,
    \end{aligned}
\end{equation}
where $\mathcal{F}_{\text{auc}}(\cdot)$ and $\mathcal{F}_{\text{acc}}(\cdot)$ are computed between the predicted responses (based on $\hat{\boldsymbol{\theta}}_i'$ and the true labels in $\mathcal{D}_i^{\text{test}}$). During this process, the model parameters are restored after evaluation to preserve consistency across candidates.

Let $\mathcal{P}_i^g = \{\mathcal{X}_i^{(1)}, \ldots, \mathcal{X}_i^{(|\mathcal{P}|)}\}$ denote the population of student $s_i$ at generation $g$, with corresponding fitness values $\mathcal{F}_i^g = [f_i^1, \ldots, f_i^{|\mathcal{P}|}]$. We sort the individuals in descending order of fitness and retain the top-$k$ elites as:
\begin{equation}
    \label{eq_12}
    \begin{aligned}
        \mathcal{E}_g^1 = \{\mathcal{X}_i^{(j)} \in \mathcal{P}_i^g \mid \text{rank}(f_i^j) \leq k \}, \quad k = \left\lfloor \frac{|\mathcal{P}|}{2} \right\rfloor.
    \end{aligned}
\end{equation}

To preserve diversity, the remaining individuals are selected by filtering the rest of the population based on Hamming distance. Specifically, each candidate is encoded as a binary bit-string $\mathbf{b}_i^j = \text{Pack}(\mathcal{X}_i^{(j)})$ and compared against the elite pool $\mathcal{B}_{\mathcal{E}}$ via batch Hamming distance:
\begin{equation}
    \label{eq_13}
    \begin{aligned}
        \text{HamDist}(\mathbf{b}_i^j, \mathcal{B}_{\mathcal{E}}) = \min_{\mathbf{b}_i^e \in \mathcal{B}_{\mathcal{E}}} \text{Hamming}(\mathbf{b}_i^j, \mathbf{b}_i^e),
    \end{aligned}
\end{equation}
where only those candidates satisfying $\text{HamDist} > \tau$ are admitted to the survivor set, and $\tau$ is a threshold (e.g., $\tau = 0.15L$). This filtering is repeated until the survivor set reaches the desired size, or a maximum number of attempts is reached. The final population is formed as:
\begin{equation}
    \label{eq_14}
    \begin{aligned}
        \mathcal{P}_i^{g+1} = \mathcal{E}_g^1 \cup \mathcal{E}_g^2, \quad \text{s.t. } |\mathcal{P}_i^{g+1}| = |\mathcal{P}|,
    \end{aligned}
\end{equation}
where $\mathcal{E}_i^g$ contains the diversity-preserved candidates sampled under the Hamming constraints.


\begin{table}[!t]
    \small
    \centering
    \label{table.dataset_statistic}
    \resizebox{0.96\linewidth}{!}{ 
        \begin{tabular}{l|rrr}
        \toprule
        Dataset & JUNYI & PTADisc \\
        \midrule
        \#Learners & 54,564 & 18,768  \\
        \#Exercises & 565& 3,262  \\
        \#Knowledge concepts & 30 & 50\\
        \#Interactions & 1,711,210 & 5,720,582 \\
        Avg. interactions per learner & 31.36 & 304.80 \\
        Avg. exercises per concept & 18.83 & 70.06\\
        \bottomrule
        \end{tabular}
    }
    \vspace{-2mm}
    \caption{The statistics of all datasets.}
    \vspace{-6mm}
\end{table}

\section{Experiments}

\subsection{Experimental Setting}
\subsubsection{Datasets.} We conducted experiments on two real-world educational datasets of different scales and characteristics, JUNYI~\cite{JUNYI} and PTADisc~\cite{PTADisc}, to evaluate the effectiveness of the proposed PEOAT on the one-shot adaptive testing~(OAT) task. The statistical overview of both datasets is presented in Table~1.

\subsubsection{Baseline Approaches.}
To demonstrate the effectiveness of the proposed model, we compare it with a comprehensive set of computerized adaptive testing approaches, including both heuristic and data-driven methods. In total, eight CAT algorithms are considered: RAND, MKLI~\cite{MKLI}, MAAT~\cite{MAAT}, BECAT~\cite{BECAT}, BOBCAT~\cite{BOBCAT}, NCAT~\cite{NCAT}, GMOCAT~\cite{GMOCAT}, and UATS~\cite{UATS}. 


\subsubsection{Evaluation Metrics.}
The goal of the OAT task is to maximize the quality of ability assessment. Following the evaluation protocol commonly used in traditional CAT settings, we adopt two standard metrics to assess model performance: the area under the ROC curve (AUC) and accuracy (ACC).

\subsubsection{Experimental Settings.}
 
In our experiment, we adopt MIRT~\cite{MIRT} and NCD~\cite{NCD} as the backbone diagnosis models of the ability estimation module. During the pre-training of $\mathcal{M}_d$, the student and item embeddings are initialized with dimensions equal to the number of knowledge concepts. In the OAT evaluation phase, the question selection model $\mathcal{M}_\pi$ adopts consistent settings, where the learning rates for MIRT and NCD updates are set to 0.02 and 0.005, respectively, with $5*\sqrt{L}$ epochs. The one-shot selection lengths $L$ are set \{5, 10, 15, 20\}. We used a population size of 20, 15 evolutionary generations, a crossover rate of 0.8, a mutation rate of 0.2, and search the distance threshold $\tau$ in \{0.5, 0.75, 1, 1.25, 1.5\}. All models are Xavier-initialized and optimized using Adam in PyTorch, with experiments conducted on two NVIDIA RTX 4090 GPUs.

\begin{table*}[!t]
    \setlength{\tabcolsep}{3pt}
    \renewcommand{\arraystretch}{1.3}
    \centering

    \resizebox{1.0\linewidth}{!}{ 
        \begin{tabular}{cc|cccc|cccc}
            \Xhline{1.2pt}
            \rowcolor{my_purple}
            \multicolumn{2}{c|}{\textbf{CDM}} & \multicolumn{4}{c|}{\textbf{MIRT}} & \multicolumn{4}{c}{\textbf{NCD}}  \\
            \Xhline{1pt}
            \multicolumn{2}{c|}{\textbf{Dataset/Metric}} & \multicolumn{8}{c}{\textbf{JUNYI / ACC/AUC(\%)} $\uparrow$} \\
            \midrule
            
            \makecell[c]{\textbf{Type}} & \multicolumn{1}{|c|}{\textbf{Methods}} & length=5  & length=10 & length=15 & length=20 & length=5  & length=10 & length=15 & length=20  \\
            \midrule
            \multirow{4}{*}{\rotatebox{90}{\textbf{Heuristic}}} 
            & \multicolumn{1}{|c|}{RNAD} & \cellcolor{my_gray_light}$67.98/68.24$ & \cellcolor{my_gray_light}$74.48/73.64$ & \cellcolor{my_gray_light}$79.60/77.73$ & \cellcolor{my_gray_light}$82.47/80.48$ & \cellcolor{my_gray_light}$67.19/79.02$ & \cellcolor{my_gray_light}$69.38/80.12$ & \cellcolor{my_gray_light}$71.41/81.08$ & \cellcolor{my_gray_light}$73.21/81.94$ \\
            & \multicolumn{1}{|c|}{MKLI} & $70.14/70.27$ & $78.03/76.64$ & $83.26/81.39$ & $86.07/84.27$ & $67.12/80.45$ & $68.39/81.29$ & $70.63/82.34$ & $72.59/83.26$ \\
            & \multicolumn{1}{|c|}{MAAT} & \cellcolor{my_gray_light}$68.45/69.50$ & \cellcolor{my_gray_light}$74.55/73.04$ & \cellcolor{my_gray_light}$77.94/75.23$ & \cellcolor{my_gray_light}$79.99/77.01$ & \cellcolor{my_gray_light}$68.66/80.36$ & \cellcolor{my_gray_light}$70.31/80.84$ & \cellcolor{my_gray_light}$72.37/81.40$ & \cellcolor{my_gray_light}$74.41/82.03$ \\
            & \multicolumn{1}{|c|}{BECAT} & $67.85/69.24$ & $74.69/73.72$ & $79.92/77.59$ & $83.80/80.90$ & $67.71/80.33$ & $68.91/81.06$ & $69.73/81.60$ & $70.73/82.02$ \\
            \midrule
            \multirow{4}{*}{\rotatebox{90}{\textbf{Data-Driven}}} & \multicolumn{1}{|c|}{BOBCAT} & \cellcolor{my_gray_light}$69.15/71.86$ & \cellcolor{my_gray_light}$77.05/\underline{77.60}$ & \cellcolor{my_gray_light}$81.66/81.12$ & \cellcolor{my_gray_light}$84.29/83.43$ & \cellcolor{my_gray_light}$\underline{70.98}/\underline{81.62}$ & \cellcolor{my_gray_light}$72.84/82.68$ & \cellcolor{my_gray_light}$74.51/83.53$ & \cellcolor{my_gray_light}$76.01/84.35$ \\
            & \multicolumn{1}{|c|}{NCAT} & $\underline{71.19}/73.48$ & $80.23/77.37$ & $82.69/\underline{81.43}$ & $\underline{84.93}/84.04$ & $70.71/80.95$ & $73.18/82.82$ & $74.20/83.46$ & $76.77/\underline{84.59}$ \\
            & \multicolumn{1}{|c|}{GMOCAT} & \cellcolor{my_gray_light}$71.47/73.19$ & \cellcolor{my_gray_light}$79.15/77.53$ & \cellcolor{my_gray_light}$82.38/80.49$ & \cellcolor{my_gray_light}$84.74/83.58$ & \cellcolor{my_gray_light}$70.55/80.86$ & \cellcolor{my_gray_light}$72.47/81.73$ & \cellcolor{my_gray_light}$\underline{74.69}/\underline{83.91}$ & \cellcolor{my_gray_light}$76.58/84.04$ \\
            &\multicolumn{1}{|c|}{UATS} & $70.83/\underline{74.45}$ & $\underline{80.33}/77.19$ & $\underline{83.13}/81.27$ & $84.38/\underline{84.65}$ & $70.44/80.61$ & $\underline{73.35}/\underline{83.16}$ & $74.18/82.87$ & $\underline{77.52}/84.15$ \\
            \midrule
            \multirow{1}{*}{\textbf{Ours}} & \multicolumn{1}{|c|}{PEOAT} & \cellcolor{my_gray}$\bm{79.64}/\bm{83.05}$ & \cellcolor{my_gray}$\bm{85.38}/\bm{85.85}$ & \cellcolor{my_gray}$\bm{86.39}/\bm{86.68}$ & \cellcolor{my_gray}$\bm{86.85}/\bm{87.83}$ & \cellcolor{my_gray}$\bm{74.56}/\bm{83.06}$ & \cellcolor{my_gray}$\bm{81.90}/\bm{86.47}$ & \cellcolor{my_gray}$\bm{85.85}/\bm{88.86}$ & \cellcolor{my_gray}$\bm{87.34}/\bm{89.78}$ \\

            \midrule
            \multicolumn{2}{c|}{\textbf{Dataset/Metric}} & \multicolumn{8}{c}{\textbf{PTADisc / ACC/AUC(\%)} $\uparrow$} \\
            \midrule
            
            \makecell[c]{\textbf{Type}} & \multicolumn{1}{|c|}{\textbf{Methods}} & length=5  & length=10 & length=15 & length=20 & length=5  & length=10 & length=15 & length=20  \\
            \midrule
            \multirow{4}{*}{\rotatebox{90}{\textbf{Heuristic}}} 
            & \multicolumn{1}{|c|}{RAND} & \cellcolor{my_gray_light}$61.29/62.93$ & \cellcolor{my_gray_light}$63.89/63.89$ & \cellcolor{my_gray_light}$65.93/64.78$ & \cellcolor{my_gray_light}$66.74/65.52$ & \cellcolor{my_gray_light}$63.83/66.71$ & \cellcolor{my_gray_light}$64.57/67.11$ & \cellcolor{my_gray_light}$65.24/67.47$ & \cellcolor{my_gray_light}$65.92/67.84$ \\
            & \multicolumn{1}{|c|}{MKLI} & $62.14/65.44$ & $64.09/66.73$ & $67.23/\underline{68.24}$ & $65.84/67.65$ & $64.95/68.41$ & $64.67/68.51$ & $64.73/68.69$ & $65.16/68.87$ \\
            & \multicolumn{1}{|c|}{MAAT} & \cellcolor{my_gray_light}$61.78/63.29$ & \cellcolor{my_gray_light}$62.90/63.84$ & \cellcolor{my_gray_light}$65.51/64.25$ & \cellcolor{my_gray_light}$65.77/65.63$ & \cellcolor{my_gray_light}$64.77/68.42$ & \cellcolor{my_gray_light}$64.71/68.65$ & \cellcolor{my_gray_light}$64.94/69.01$ & \cellcolor{my_gray_light}$65.11/69.32$ \\
            & \multicolumn{1}{|c|}{BECAT} & $62.34/64.80$ & $63.43/64.85$ & $64.74/65.12$ & $66.10/65.43$ & $65.25/68.42$ & $65.79/68.94$ & $66.64/69.38$ & $67.18/69.52$ \\
            \midrule
            \multirow{4}{*}{\rotatebox{90}{\textbf{Data-Driven}}} & \multicolumn{1}{|c|}{BOBCAT} & \cellcolor{my_gray_light}$62.87/65.04$ & \cellcolor{my_gray_light}$\underline{65.32}/66.17$ & \cellcolor{my_gray_light}$67.32/67.10$ & \cellcolor{my_gray_light}$69.01/67.81$ & \cellcolor{my_gray_light}$65.95/\underline{68.71}$ & \cellcolor{my_gray_light}$66.70/69.13$ & \cellcolor{my_gray_light}$67.39/69.52$ & \cellcolor{my_gray_light}$68.08/69.94$ \\
            & \multicolumn{1}{|c|}{NCAT} & $63.19/65.23$ & $64.82/\underline{67.45}$ & $67.96/67.87$ & $69.55/68.39$ & $66.38/68.09$ & $\underline{67.64}/\underline{69.22}$ & $68.48/\underline{69.61}$ & $\underline{70.17}/\underline{70.40}$ \\
            & \multicolumn{1}{|c|}{GMOCAT} & \cellcolor{my_gray_light}$\underline{63.73}/\underline{65.80}$ & \cellcolor{my_gray_light}$65.11/66.86$ & \cellcolor{my_gray_light}$67.52/67.27$ & \cellcolor{my_gray_light}$69.61/68.05$ & \cellcolor{my_gray_light}$\underline{66.47}/68.24$ & \cellcolor{my_gray_light}$67.48/68.97$ & \cellcolor{my_gray_light}$\underline{69.14}/69.49$ & \cellcolor{my_gray_light}$69.73/70.36$ \\
            &\multicolumn{1}{|c|}{UATS}  & $63.51/64.94$ & $64.90/67.29$ & $\underline{68.23}/67.51$ & $\underline{70.25}/\underline{68.67}$ & 
            $66.09/68.33$ &$67.47/68.86$ & $68.63/69.35$ & $69.81/70.03$ \\
            \midrule
            \multirow{1}{*}{\textbf{Ours}} & \multicolumn{1}{|c|}{PEOAT} & \cellcolor{my_gray}$\bm{68.10}/\bm{69.80}$ & \cellcolor{my_gray}$\bm{71.96}/\bm{71.59}$ & \cellcolor{my_gray}$\bm{73.05}/\bm{72.36}$ & \cellcolor{my_gray}$\bm{74.17}/\bm{72.65}$ & \cellcolor{my_gray}$\bm{69.37}/\bm{70.84}$ & \cellcolor{my_gray}$\bm{73.65}/\bm{73.58}$ & \cellcolor{my_gray}$\bm{75.44}/\bm{75.07}$ & \cellcolor{my_gray}$\bm{75.91}/\bm{74.93}$ \\
            
            \bottomrule
        \end{tabular}

    }
    \caption{Performance comparison of PEOAT and baselines on the JUNYI and PTADisc datasets in terms of ACC and AUC. \textbf{Bold} highlights the best performance (statistically significant at $p<0.05$), and \underline{underline} marks the second-best performance.}
    \vspace{-1mm}
    \label{table.Result_Table}
\end{table*}

\subsection{Performance Comparison}

Table~2 presents the experimental results of the proposed PEOAT model for one-shot adaptive testing, compared with all baseline methods on the two datasets. The best performance for each metric is highlighted in bold, while the second-best is underlined. According to the results, there are several observations: 
\textbf{(1)}~PEOAT consistently outperforms all state-of-the-art baselines across both datasets and question lengths. Specifically, compared to the second-best model, it achieves average improvements of 7.74\% and 5.82\% in ACC and AUC on the JUNYI dataset, and 6.97\% and 5.62\% in ACC and AUC on the PTADisc dataset, respectively. This consistent advantage suggests that PEOAT’s personalization-guided selection effectively aligns exercise assembly with individual diagnostic objectives;~\textbf{(2)}~The superiority of PEOAT is particularly pronounced at shorter testing lengths. For instance, on the JUNYI dataset under the MIRT diagnosis model, PEOAT outperforms the second-best baseline by 10.61\% in ACC and 10.35\% in AUC at $length = 5$, and by 3.77\% and 6.05\% in ACC and AUC at $length = 15$, respectively. These results further highlight the strong potential and practical applicability of PEOAT in fast, one-shot question assembly scenarios. In addition, we compared the performance of the basic version of PEOAT without targeted design, as presented in Table~3. The results demonstrate that formulating the OAT task as a combinatorial optimization problem and incorporating the evolutionary algorithm significantly enhance performance, further validating the superiority of the proposed PEOAT model.

\subsection{Ablation Study}
We conducted a comprehensive ablation study to investigate the contribution of each module in the PEOAT framework by defining the following variants: 1) \textbf{w/o PI}: removing the personalization-aware population initialization and replacing it with random initialization only; 2) \textbf{w/o CE}: removing the cognitive-enhanced evolutionary search strategy and replacing it with basic crossover and mutation operations; 3) \textbf{w/o ES}: removing the diversity-preserving environmental selection. To conserve space, we provide the accuracy results of MIRT as a basic dianosis model on the JUNYI dataset. As illustrated in Figure~3, the results reveal insightful observations: (1) Compared to PEOAT, all variants exhibit relative performance degradation, highlighting the contribution of the designed sub-modules to our proposed model. (2) The most significant performance drop occurs when the population initialization strategy is removed, indicating that the incorporation of personalized information substantially enhances the quality of the initial population.

\begin{table}[!t]
    \setlength{\tabcolsep}{4pt}
    \renewcommand{\arraystretch}{1.3}
    \centering
    \label{table.Result_Table}

    \resizebox{1.0\linewidth}{!}{ 
        \begin{tabular}{cc|cccc}
            \Xhline{1.2pt}
            \midrule
            \multicolumn{2}{c|}{\textbf{Dataset/Metric}} & \multicolumn{4}{c}{\textbf{JUNYI / ACC/AUC(\%)} $\uparrow$} \\
            \midrule
            
            \makecell[c]{\textbf{CDM}} & \multicolumn{1}{|c|}{\textbf{Methods}} & length=5  & length=10 & length=15 & length=20 \\
            
            \midrule
            \multirow{2}{*}{\rotatebox{90}{\textbf{MIRT}}} & \multicolumn{1}{|c|}{PEOAT-B} & $78.35/81.97$ & $84.12/84.48$ & $84.96/85.21$ & $85.73/86.55$  \\
            & \multicolumn{1}{|c|}{PEOAT} & $\bm{79.64}/\bm{83.05}$ & $\bm{85.38}/\bm{85.85}$ & $\bm{86.39}/\bm{86.68}$ & $\bm{86.85}/\bm{87.83}$ \\
            \midrule
            \multirow{2}{*}{\rotatebox{90}{\textbf{NCD}}} & \multicolumn{1}{|c|}{PEOAT-B} & $73.27/81.81$ & $80.64/85.19$ & $84.73/87.59$ & $86.11/88.80$  \\
            & \multicolumn{1}{|c|}{PEOAT} & $\bm{74.56}/\bm{83.06}$ & $\bm{81.90}/\bm{86.47}$ & $\bm{85.85}/\bm{88.86}$ & $\bm{87.34}/\bm{89.78}$  \\
            \midrule
            \multicolumn{2}{c|}{\textbf{Dataset/Metric}} & \multicolumn{4}{c}{\textbf{PTADisc / ACC/AUC(\%)} $\uparrow$} \\
            \midrule
            
            \makecell[c]{\textbf{CDM}} & \multicolumn{1}{|c|}{\textbf{Methods}} & length=5  & length=10 & length=15 & length=20 \\
            
            \midrule
            \multirow{2}{*}{\rotatebox{90}{\textbf{MIRT}}} & \multicolumn{1}{|c|}{PEOAT-B} & $66.57/68.24$ & $70.44/70.25$ & $71.71/71.08$ & $72.86/71.30$  \\
            & \multicolumn{1}{|c|}{PEOAT} & $\bm{68.10}/\bm{69.80}$ & $\bm{71.96}/\bm{71.59}$ & $\bm{73.05}/\bm{72.36}$ & $\bm{74.17}/\bm{72.65}$ \\
            \midrule
            \multirow{2}{*}{\rotatebox{90}{\textbf{NCD}}} & \multicolumn{1}{|c|}{PEOAT-B} & $67.93/69.41$ & $72.26/72.09$ & $74.67/73.88$ & $74.79/73.53$  \\
            & \multicolumn{1}{|c|}{PEOAT} & $\bm{69.37}/\bm{70.84}$ & $\bm{73.65}/\bm{73.58}$ & $\bm{75.44}/\bm{75.07}$ & $\bm{75.91}/\bm{74.93}$  \\
            
            \bottomrule
        \end{tabular}

    }
    \caption{Performance comparison of PEOAT and its base version PEOAT-B on the JUNYI and PTADisc datasets.}
    \vspace{-1mm}
\end{table}

\begin{figure}[!t] 
    \vspace{-3mm}
	\centering 
	\includegraphics[scale=0.47]{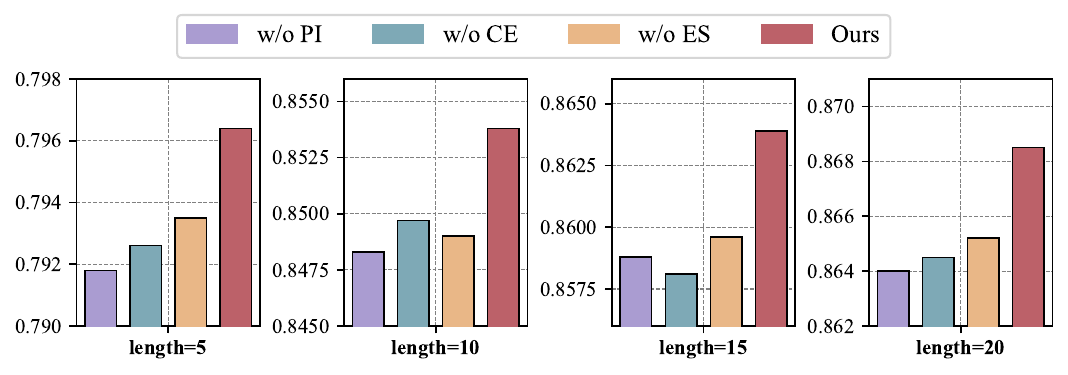} 
    \vspace{-5mm}
	\caption{Performance of ablation studies conducted on the JUNYI, where “w/o” means removing the target module.}
    \label{fig.ablition} 
\end{figure}

\subsection{Parameter Sensitivity Analysis}
 
In this section, we conducted a parameter sensitivity analysis to examine the impact of key hyper-parameters, with a primary focus on the distance threshold $\tau$ used in the diversity-preserving environmental selection. Specifically, we set $\tau$ to \{0.5, 0.75, 1.0, 1.25, 1.5\}, and primarily report the experimental results on the JUNYI dataset. As shown in Figure~4, the model achieves its best performance when $\tau$ is set to 1.0, under testing lengths of 10 and 20. Notably, as the threshold varies, the model’s performance does not exhibit a strictly consistent pattern or a clear linear trend. Nevertheless, the overall tendency roughly follows an initial increase followed by a decrease, which may be impacted by the testing length.

\subsection{Case Study}
 
To further investigate the evolution of question populations and the convergence of search strategies in PEOAT’s question selection, we conduct two case studies in this section. Specifically, 20 students with similar ability levels from the JUNYI dataset are selected, and their ability estimation performance (accuracy and fitness) is tracked during population evolution under varying test lengths, using MIRT as the base model. Figure~5 presents the performance evolution with error bands under two metrics. It can be observed that the assessment performance of individual students improves significantly as the population evolves across different test lengths, particularly in terms of fitness, highlighting the effectiveness of PEOA in evolutionary search. Meanwhile, we also sampled two student groups and visualized the evolution of their overall assessment performance using cloud-rain plots. As shown in Figure~6, both groups exhibit an upward performance trend under test lengths of 5 and 10, gradually converging as the number of generations increases. This indicates that student performance not only improved but also became more consistent over time.

\begin{figure}[!t] 
    \vspace{-2mm}
	\centering 
	\includegraphics[scale=0.37]{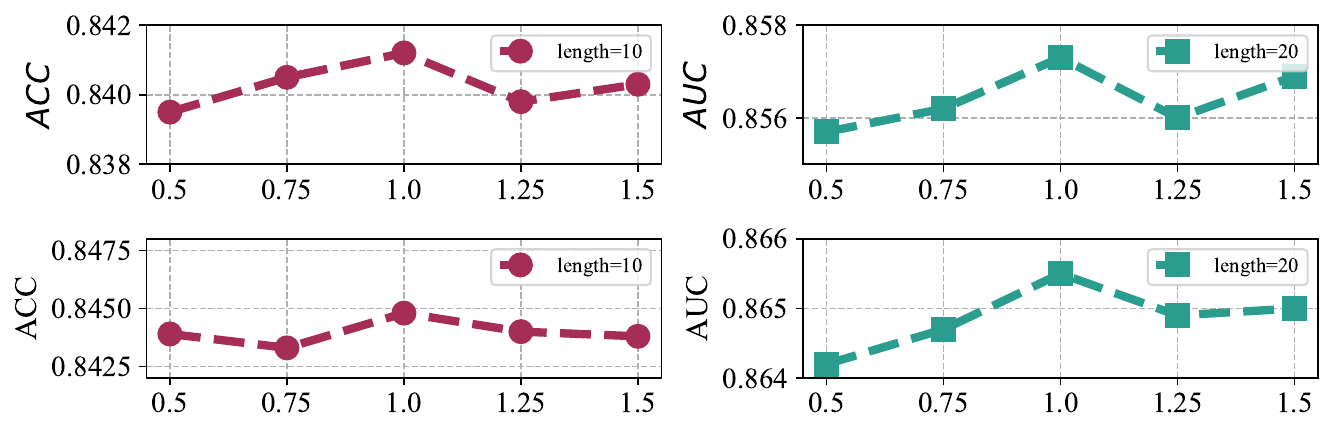} 
	\caption{Sensitivity analysis of the distance threshold $\tau$ of the environmental selection on the JUNYI dataset.}
    \label{fig.para_study_2} 
    \vspace{-3mm}
\end{figure}

\begin{figure}[!t] 
	\centering 
	\includegraphics[scale=0.90]{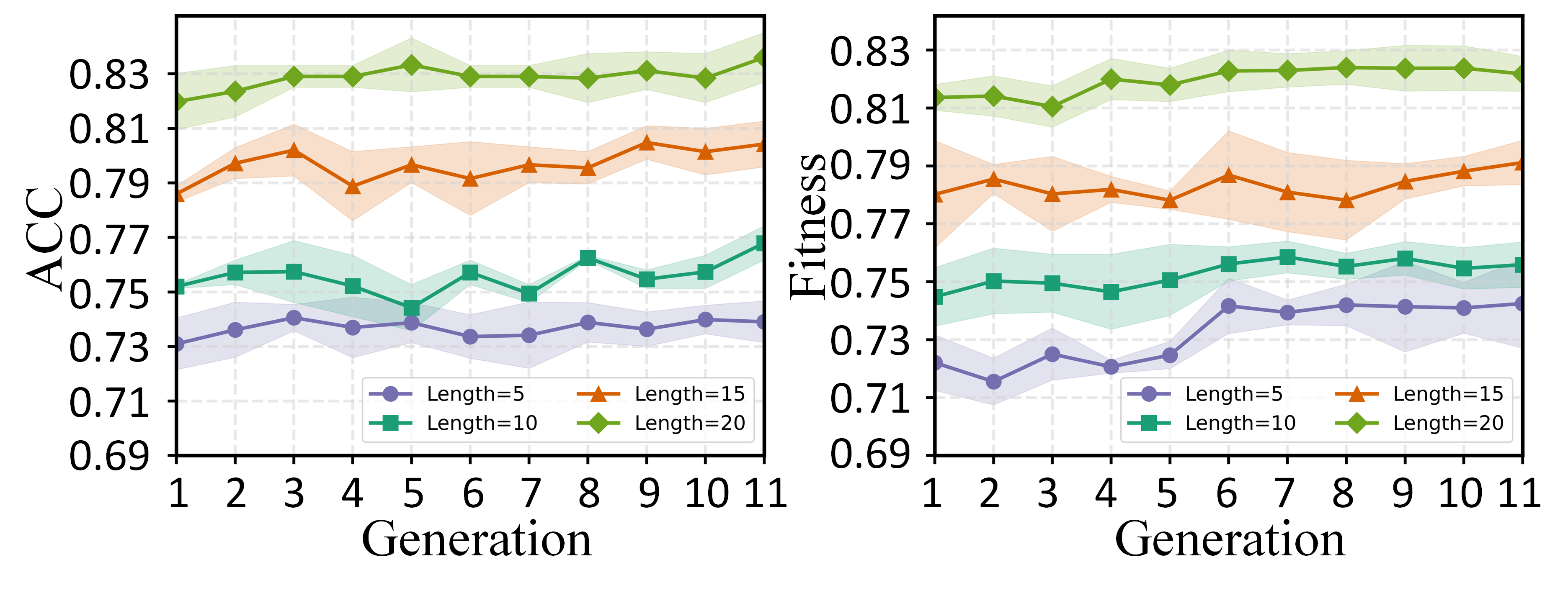} 
    \vspace{-6mm}
	\caption{Case study of the performance evolution of the assembled question populations on the JUNYI dataset.}
    \label{fig.case_study_1} 
\end{figure}

\begin{figure}[!t] 
	\centering 
	\includegraphics[scale=0.84]{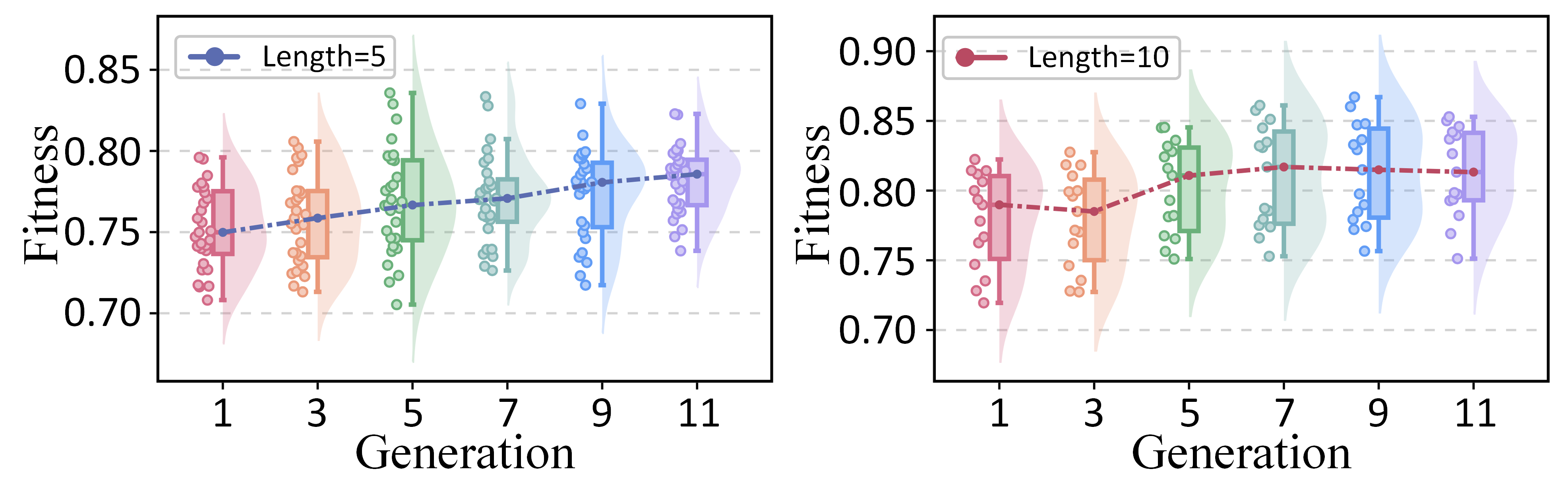} 
    \vspace{-6mm}
	\caption{Case study of the generational fitness progression across varying test lengths on the JUNYI dataset.}
    \vspace{-3mm}
    \label{fig.case_study_2} 
\end{figure}

\section{Conclusion}

In this paper, we first proposed a novel task called One-Shot Adaptive Testing (OAT). This task posed three major challenges: ensuring student adaptability during optimization, effectively searching an enormous solution space, and alleviating encoding sparsity due to a candidate pool far exceeding test length. To address these, we introduced \textbf{PEOAT}, a \underline{\textbf{P}}ersonalization-guided \underline{\textbf{E}}volutionary question assembly framework for \underline{\textbf{O}}ne-shot \underline{\textbf{A}}daptive \underline{\textbf{T}}esting. We first designed a personalization-aware population initialization method that incorporated individual ability and exercise difficulty, using multi-strategy sampling to build a diverse and effective initial search space. Then, we developed a cognitive-enhanced evolutionary search incorporating schema-preserving crossover and cognitive information-guided mutation operators to enable efficient exploration. Finally, a diversity-preserving environmental selection strategy was implemented to maintain population diversity while considering fitness. Extensive experiments on two real educational datasets demonstrated the model’s effectiveness, and additional case studies provided valuable insights.

\section{Acknowledgements}

This work was supported in part by the National Natural Science Foundation of China (No. U21A20512, No.62107001, No.62302010), in part by the Anhui Province Key Laboratory of Intelligent Computing and Applications (No. AFZNJS2024KF01), and in part by the Anhui Provincial Natural Science Foundation(No. 2508085MF160).

\bibliography{references}

\end{document}